\documentclass[11pt]{article}
\usepackage{times}
\usepackage{mathfont}
\usepackage{graphicx}
\usepackage{cite}
\usepackage{url}
\mathcode`O="724F

\newtheorem{theorem}{Theorem}

\newcommand{\qed}{~$\vrule width.15cm height.2cm depth0cm$ \medbreak}
\newenvironment{proof}{\noindent{\bf Proof: }}{\qed}

\begin{document}

\title{Phutball Endgames are Hard}

\author{Erik D. Demaine\thanks{University of Waterloo,
Department of Computer Science,
Waterloo, Ontario N2L 3G1, Canada,
{\tt \{eddemaine,mldemaine\}@uwaterloo.ca}}
\and
Martin L. Demaine\footnotemark[1]
\and
David Eppstein\thanks{University of California, Irvine,
Department of Information and Computer Science,
Irvine, CA 92697-3425, USA,
{\tt eppstein@ics.uci.edu}}}

\date{ }
\maketitle

\begin{abstract}
We show that, in John Conway's board game Phutball (or Philosopher's
Football), it is NP-complete to determine whether the current player has
a move that immediately wins the game.  In contrast, the similar
problems of determining whether there is an immediately winning move in
checkers, or a move that kings a man, are both solvable in
polynomial time.
\end{abstract}

\section{Introduction}

\begin{figure}[t]
\begin{center}
\includegraphics[width=3in]{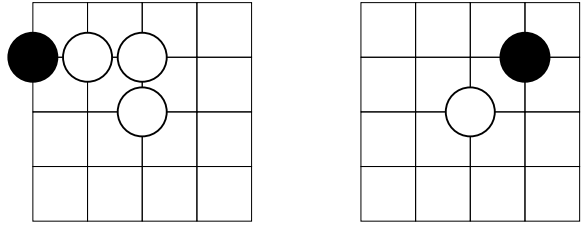}
\end{center}
\caption{A jump in Phutball. Left: The situation prior to the jump.
Right: The situation after jumping two men. The same player may then
jump the remaining man.}
\label{fig:jump}
\end{figure}

John Conway's game Phutball \cite{BerConGuy-82,Bur-99,Caz-99,Wil-AG-00},
also known as Philosopher's Football, starts with a single black stone
(the {\em ball}) placed at the center intersection of a rectangular grid
such as a Go board.  Two players sit on opposite sides of the board and
take turns.  On each turn, a player may either place a
single white stone (a {\em man}) on any vacant intersection, or
perform a sequence of {\em jumps}.  To jump, the ball must be adjacent
to one or more men.  It is moved in a straight line (orthogonal or
diagonal) to the first vacant intersection beyond the men, and the men so
jumped are immediately removed (Figure~\ref{fig:jump}).  If a jump is
performed, the same player may continue jumping as long as the ball
continues to be adjacent to at least one man, or may end the turn at any
point.  Jumps are not obligatory: one can choose to place a man instead
of jumping.  The game is over when a jump sequence ends on or over the
edge of the board closest to the opponent (the opponent's {\em goal
line}) at which point the player who performed the jumps wins.  It is
legal for a jump sequence to step onto but not over one's own goal line. 
One of the interesting properties of Phutball is that any move could be
played by either player, the only partiality in the game being the rule
for determining the winner.

It is theoretically possible for a Phutball game to return to a previous
position, so it may be necessary to add a loop-avoidance rule such as the
one in Chess (three repetitions allow a player to claim a draw) or Go
(certain repeated positions are disallowed).  However, repetitions
do not seem to come up much in actual practice.

It is common in other board games\footnote{More precisely, since most
games have a finite prescribed board size, these complexity results apply
to generalizations in which arbitrarily large boards are
allowed, and in which the complexity is measured in terms of the
board size.} that the problems of determining the outcome of the game
(with optimal play), or testing whether a given move preserves the correct
outcome, are PSPACE-complete
\cite{Epp-00}, or even EXPTIME-complete for loopy games such as Chess
\cite{FraLic-JCT-81} and Go \cite{Rob-IFIP-83}.  However, no such result is
known for Phutball.  Here we prove a different kind of complexity result:
the problem of determining whether a player has a move that immediately
wins the game (a mate in one, in chess terminology) is NP-complete.  Such
a result seems quite unusual, since in most games there are only a small
number of legal moves, which could all be tested in polynomial time. The
only similar result we are aware of is that, in Twixt, it is NP-complete
to determine whether a player's points can be connected to form a winning
chain \cite{MazWat-97}. However, the Twixt result seems to be less
applicable to actual game play, since it depends on a player making a
confusing tangle of his own points, which does not tend to occur in
practice. The competition between both players in Phutball to form a
favorable arrangement of men seems to lead much more naturally to
complicated situations not unlike the ones occurring in our
NP-completeness proof.

\section{The NP-Completeness Proof}

Testing for a winning jump sequence is clearly in NP,
since a jump sequence can only be as long as the number of men on the
board.  As is standard for NP-completeness proofs, we prove that the
problem is hard for NP by reducing a known NP-complete problem to it.
For our known problem we choose 3-SAT: satisfiability of Boolean
formulae in conjunctive normal form, with at most three variables per
clause.  We must show how to translate a 3-SAT instance into a Phutball
position, in polynomial time, in such a way that the given formula is
solvable precisely if there exists a winning path in the Phutball
position.

\begin{figure}[t]
\begin{center}
\includegraphics[width=3in]{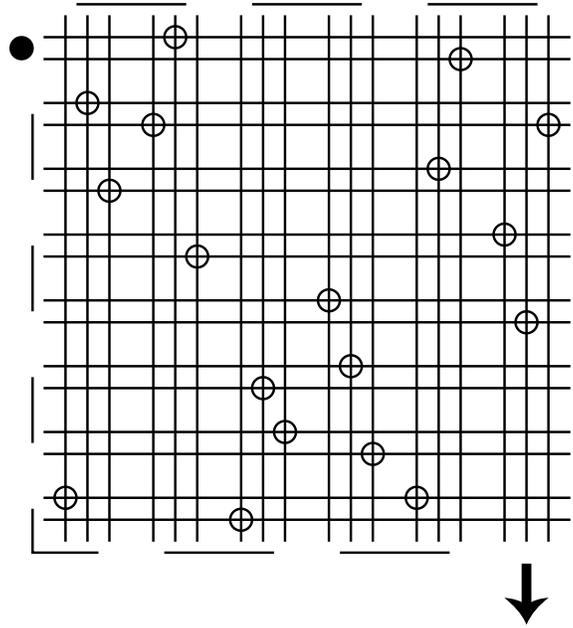}
\end{center}
\caption{Overall plan of the NP-completeness reduction: a path
zigzags through horizontal pairs of lines (representing variables)
and vertical triples of lines (representing clauses). Certain
variable-clause crossings are marked, representing an interaction
between that variable and clause.}
\label{fig:plan}
\end{figure}

\begin{figure}[p]
\begin{center}
\includegraphics[width=3in]{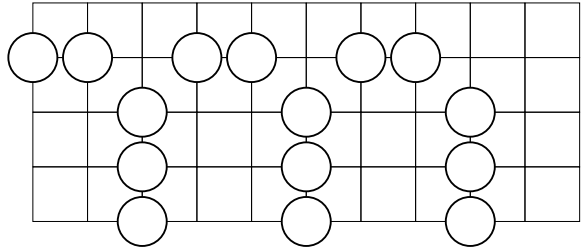}
\end{center}
\caption{Configuration of men to allow a choice between three vertical
lines. Similar configurations are used at the other end of each triple of
lines, and at each end of pairs of horizontal lines.}
\label{fig:fanout}
\end{figure}

\begin{figure}[p]
\begin{center}
\includegraphics[width=3in]{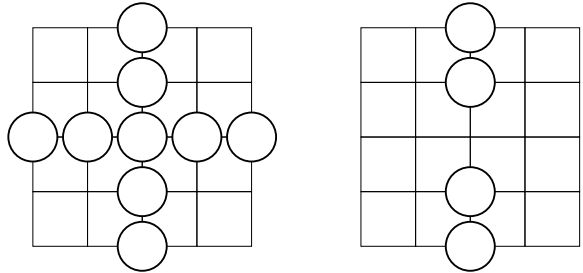}
\end{center}
\caption{Left: configuration of men to allow horizontal and vertical lines
to cross without interacting. Right: after the horizontal jump has been
taken, the short gap in the vertical line still allows it to be traversed
via a pair of jumps.}
\label{fig:cross}
\end{figure}

\begin{figure}[p]
\begin{center}
\includegraphics[width=3in]{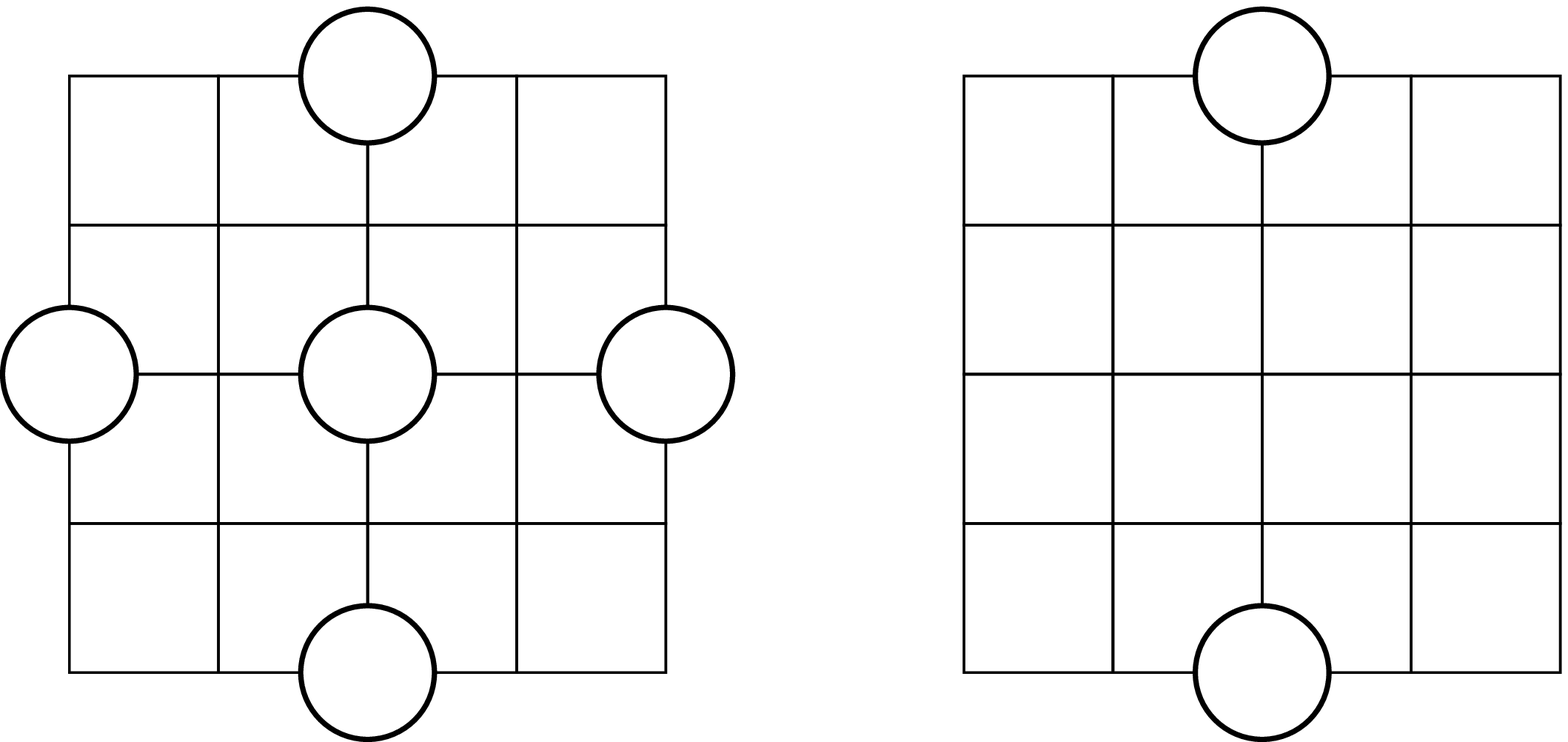}
\end{center}
\caption{Left: configuration of men to allow horizontal and vertical lines
to interact. Right: after the horizontal jump has been
taken, the long gap in the vertical line prevents passage.}
\label{fig:interact}
\end{figure}

\begin{figure}[p]
\begin{center}
\includegraphics[width=6in]{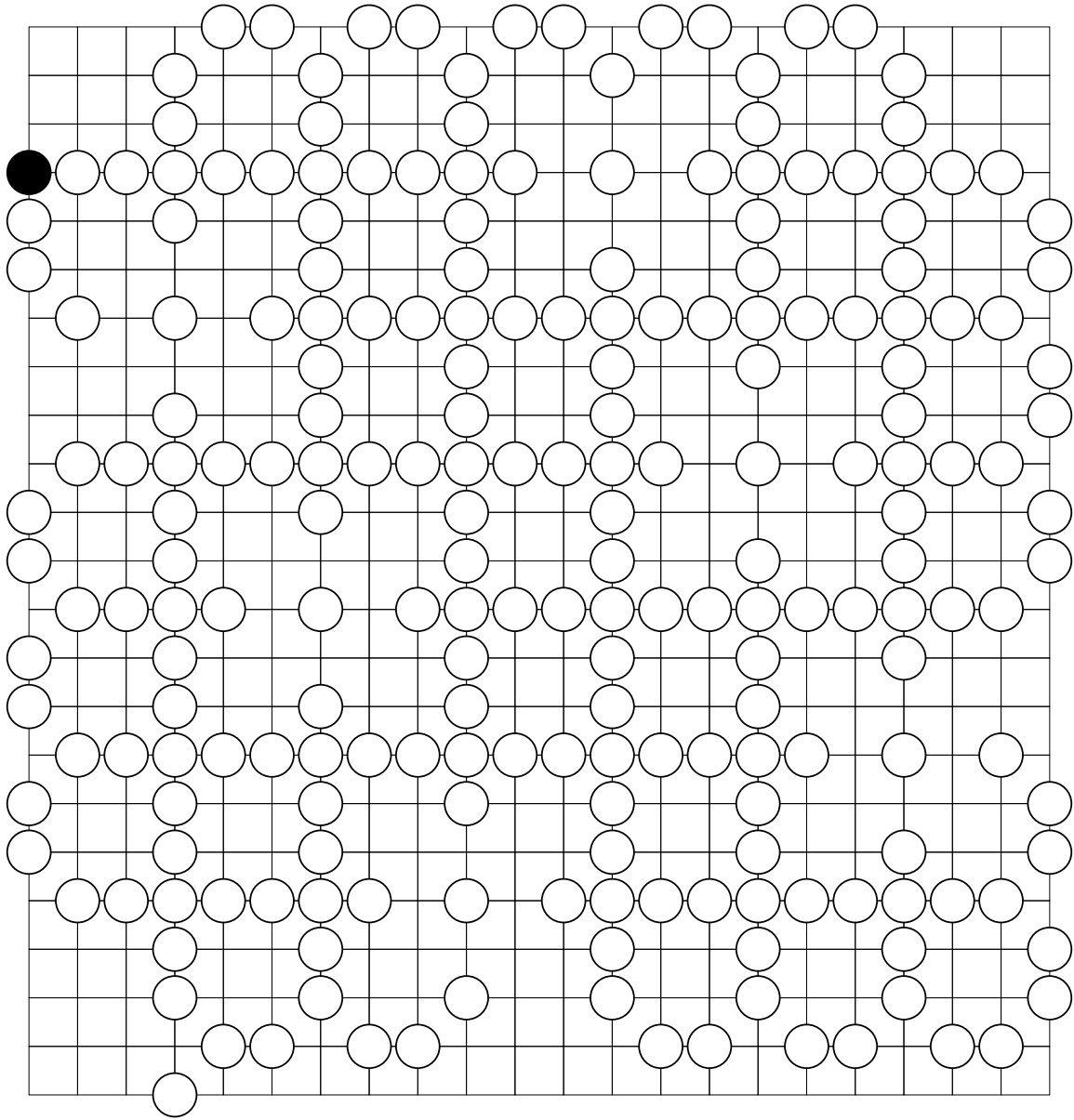}
\end{center}
\caption{Complete translation of 3-SAT instance
$(a\vee b\vee c)\wedge (\neg a\vee\neg b\vee\neg c)$.}
\label{fig:full}
\end{figure}

The overall structure of our translation
is depicted in Figure~\ref{fig:plan}, and a small complete example is
shown in Figure~\ref{fig:full}.
We form a Phutball position in which the possible jump sequences zigzag
horizontally along pairs of lines, where each pair represents one of the
variables in the given 3-SAT instance.  The path then zigzags vertically
up and down along triples of lines, where each triple represents one of
the clauses in the 3-SAT instance.  Thus, the potential winning paths
are formed by choosing one of the two horizontal lines in each pair
(corresponding to selecting a truth value for each variable)
together with one of the three vertical lines in each triple
(corresponding to selecting which of the three variables has a truth
value that satisfies the clause).  By convention, we associate paths
through the upper of a pair of horizontal lines with assignments that set
the corresponding variable to true, and paths through the lower of the
pair with assignments that set the variable to false.  The horizontal and
vertical lines interact at certain marked crossings in a way that forces
any path to correspond to a satisfying truth assignment.

We now detail each of the components of this structure.

\begin{description}
\item[Fan-in and fan-out.]
At the ends of each pair or triple of lines, we need a configuration
of men that allows paths along any member of the set of lines to converge,
and then to diverge again at the next pair or triple. 
Figure~\ref{fig:fanout} depicts such a configuration for the triples of
vertical lines; the configuration for the horizontal lines is similar.
Note that, if a jump sequence enters the configuration from the left, it
can only exit through one of the three lines at the bottom.  If a jump
sequence enters via one of the three vertical lines, it can exit
horizontally or on one of the other vertical lines.  However, the
possibility of using more than one line from a group does not cause a
problem: a jump sequence that uses the second of two horizontal lines
must get stuck at the other end of the line, and a jump sequence that
uses two of three vertical lines must use all three lines and can be
simplified to a sequence using only one of the three lines.

\item[Non-interacting line crossing.]
Figure~\ref{fig:cross} depicts a configuration of men that allows
two lines to cross without interacting. A jump sequence entering along
the horizontal or vertical line can and must exit along the same line,
whether or not the other line has already been jumped.

\item[Interaction.]
Figure~\ref{fig:interact} depicts a configuration of men forming an
interaction between two lines.  In the initial configuration, a jump
sequence may follow either the horizontal or the vertical line.
However, once the horizontal line has been jumped, it
will no longer be possible to jump the vertical line.
\end{description}

\begin{theorem}
Testing whether a Phutball position allows a winning jump sequence
is NP-complete.
\end{theorem}

\begin{proof}
As described above, we reduce 3-SAT to the given problem by forming a
configuration of men with two horizontal lines of men for each variable,
and three vertical lines for each clause.  We connect these lines by the
fan-in and fan-out gadget depicted in Figure~\ref{fig:fanout}.
If variable $i$ occurs
as the $j$th term
of clause $k$, we place an interaction gadget (Figure~\ref{fig:interact})
at the point where the bottom horizontal line
in the $i$th pair of horizontal lines crosses the $j$th vertical line in
the $k$th triple of vertical lines.
If instead the negation of variable $i$ occurs in clause $k$, we place an
interaction gadget similarly, but on the top horizontal line in the pair.
At all other crossings of horizontal
and vertical lines we place the crossing gadget depicted in
Figure~\ref{fig:cross}.  Finally, we form a path of men from the final
fan-in gadget (the arrow in Figure~\ref{fig:plan}) to the goal line of
the Phutball board.

The lines from any two adjacent interaction gadgets must be separated by
four or more units, but other crossing types allow a three-unit
separation.  By choosing the order of the variables in each clause, we
can make sure that the first variable differs from the last variable of
the previous clause, avoiding any adjacencies between interaction
gadgets. Thus, we can space all lines three units apart.  If the
3-SAT instance has
$n$ variables and
$m$ clauses, the resulting Phutball board requires $6n+O(1)$ rows and
$9m+O(1)$ columns, polynomial in the input size.

Finally, we must verify that the 3-SAT instance is solvable precisely
if the Phutball instance has a winning jump sequence. Suppose first that
the 3-SAT instance has a satisfying truth assignment; then we can form a
jump sequence that uses the top horizontal line for every true variable,
and the bottom line for every false variable.  If a clause is satisfied
by the $j$th of its three variables, we choose the $j$th of the three
vertical lines for that clause.  This forms a valid jump sequence:
by the assumption that the given truth assignment satisfies the formula,
the jump sequence uses at most one of every two lines in every
interaction gadget. Conversely, suppose we have a winning jump sequence
in the Phutball instance; then as discussed above it must use one of
every two horizontal lines and one or three of every triple of vertical
lines. We form a truth assignment by setting a variable to true if its
upper line is used and false if its lower line is used. This must be a
satisfying truth assignment: the vertical line used in each clause
gadget must not have had its interaction gadget crossed horizontally,
and so must correspond to a satisfying variable for the clause.
\end{proof}

Figure~\ref{fig:full} shows the complete reduction for a simple 3-SAT
instance.  We note that the Phutball instances created by this reduction
only allow orthogonal jumps, so the rule in Phutball allowing diagonal
jumps is not essential for our result.

\section{Phutball and Checkers}
 
Phutball is similar in many ways to Checkers. As in Phutball,
Checkers players sit at opposite ends of a rectangular board, move pieces
by sequences of jumps, remove jumped pieces, and attempt to move a piece
onto the side of the board nearest the opponent.  As in Phutball, the
possibility of multiple jumps allows a Checkers player to have an
exponential number of available moves.  Checkers is PSPACE-complete
\cite{FraGarJoh-FOCS-78} or EXPTIME-complete \cite{Rob-84},
depending on the termination rules,
but these results rely on the difficulty of game
tree search rather than the large number of moves available at any
position.  Does Checkers have the same sort of single-move
NP-completeness as Phutball?

\begin{figure}[t]
\begin{center}
\includegraphics[width=4in]{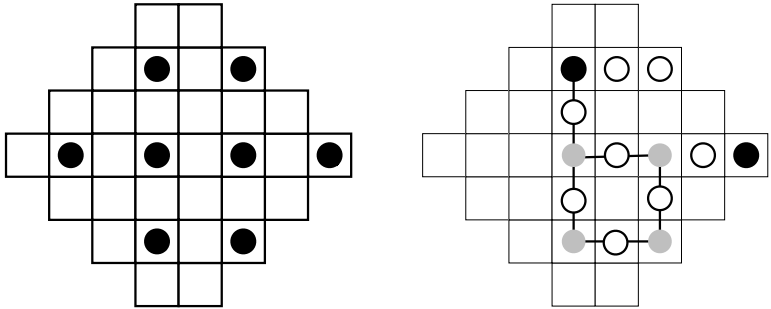}
\end{center}
\caption{Left: only cells of the same parity can be reached by jumps.
Right: graph $G_p$ formed by connecting jumpable pieces with cells that
can be reached by jumps from the upper black king.}
\label{fig:parity}
\end{figure}

It is convenient to view Checkers as being played on a nonstandard
diamond-shaped grid of square cells, with pieces that move horizontally
and vertically, rather than the usual pattern of diagonal moves on a
checkerboard (Figure~\ref{fig:checkers}).  This view does not involve
changing the rules of checkers nor even the geometric positions of the
pieces, only the markings of the board on which they rest.
Then, any jump preserves the parity of both the $x$- and $y$-coordinates
of the jumping piece, so at most one fourth of the board's cells can be
reached by jumps of a given piece (Figure~\ref{fig:parity}, left).

For any given piece $p$, form a bipartite graph $G_p$ by connecting the
vacant positions that $p$ can reach by jumping with the adjacent pieces of
the opposite color that $p$ can jump. If $p$ is a king, this graph
should be undirected, but otherwise it should be directed according to the
requirement that the piece not move backwards. Note that each jumpable
piece has degree two in this graph, so the possible sequences of jumps
are simply the graph paths that begin at the given piece and end at a
vacant square. Figure~\ref{fig:parity}, right, depicts an example; note
that opposing pieces that can not be jumped (because they are on cells
of the wrong parity, or because an adjacent cell is occupied) are not
included in $G_p$. Using this structure, it is not hard to show that
Checkers moves are not complex:

\begin{figure}[t]
\begin{center}
\includegraphics[width=4in]{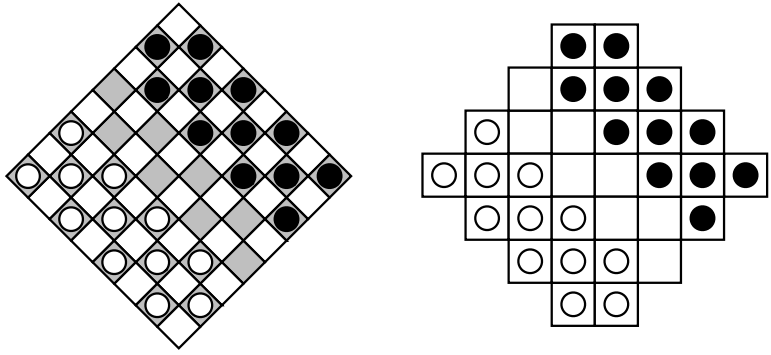}
\end{center}
\caption{The checkerboard can equivalently be viewed as a
diamond-shaped grid of orthogonally adjacent square cells.}
\label{fig:checkers}
\end{figure}

\begin{theorem}
For any Checkers position (on an arbitrary-size board), one can test in
polynomial time whether a checker can become a king, or whether there
is a move which wins the game by jumping all the opponent's pieces.
\end{theorem}

\begin{proof}
Piece $p$ can king precisely if there is a directed path in $G_p$ from
$p$ to one of the squares along the opponent's side of the board.
A winning move exists precisely if there exists a piece $p$ for which
$G_p$ includes all opposing pieces and contains an Euler path starting at
$p$; that is, precisely if
$G_p$ is connected and has at most one odd-degree vertex other than the
initial location of $p$.
\end{proof}

The second claim in this theorem, testing for a one-move win,
is also proved in \cite{FraGarJoh-FOCS-78}.  That paper also show that
the analogous problem for a generalization of checkers to arbitrary graphs
is NP-complete.

\section{Discussion}

We have shown that, in Phutball, the exponential number of jump sequences
possible in a single move, together with the ways in which parts of a
jump sequence can interfere with each other, leads to the high
computational complexity of finding a winning move.
In Checkers, there may be exponentially many jump sequences, but jumps
can be performed independently of each other, so finding winning moves is
easy.  What about other games?

In particular, Fanorona~\cite{Epp-97,Fox-87} seems a natural candidate
for study.  In this game, capturing is performed in a different way, by
moving a piece in one step towards or away from  a contiguous line of
the opponent's pieces.  Board squares alternate between strong (allowing
diagonal moves) and weak (allowing only orthogonal moves), and a piece
making a sequence of captures must change direction at each step.  Like
Checkers (and unlike Phutball) the game is won by capturing all the
opponent's pieces rather than by reaching some designated goal.  Is
finding a winning move in Fanorona hard? If so, a natural candidate
for a reduction is the problem of finding Hamiltonian paths in
grid graphs~\cite{ItaPapSzw-SJC-82}.

The complexity of determining the outcome of a general Phutball position
remains open.  We have not even proven that this problem is NP-hard, since
even if no winning move exists in the positions we construct, the player
to move may win in more than one move.

\section*{Acknowledgment}

This work was inspired by various people, including John Conway and Richard
Nowakowski, playing phutball at the MSRI Combinatorial Game Theory Research
Workshop in July 2000.

\bibliographystyle{abuser}
\bibliography{phutball}

\end{document}